\documentstyle[11pt,newpasp,twoside,epsf]{article}

\reversemarginpar

\begin{document}

\title{High Resolution 18 $\micron$ Imaging of Hot Molecular Cores}
\author{James M. De Buizer}
\affil{Cerro Tololo Inter-American Observatory, Casilla 603, La Serena,
Chile}

\begin{abstract}
I present the latest results from a search for hot molecular cores at
mid-infrared wavelengths from the largest optical telescopes available at
the present time. Three well-observed hot molecular cores were imaged,
G29.96-0.02, G19.61-0.23, and G34.26+0.15. Even though mid-infrared sources
have been claimed to be detected previously at the hot molecular core locations 
of both G19.61-0.23 and G34.26+0.15, only the hot molecular core in
G29.96-0.02 resulted in a detection. New upper limits on mid-infrared emission
are given for the hot molecular cores that were not detected.
\end{abstract}

\section{Introduction}

Hot molecular cores (HMCs) are believed to represent the earliest stages of
massive stellar birth, however there are very few observations of these
objects to date. The basic characteristics of hot molecular cores are: 1)
they are compact sources seen in radio wavelength ammonia (or molecular
line) images but have little or no radio continuum emission; 2) they lie in
massive star forming regions near ultracompact HII (UC HII) regions; 3) they
are too young and/or embedded to be seen in the optical or near infrared;
and 4) they are often coincident with water maser emission. Only a small
number of sources exist that have had such a holistic set of observations
performed.

Cesaroni et al. (1994) argue that the dust and gas in these HMCs are
well-mixed and that the gas temperatures derived from ammonia line imaging
should be an approximation to the dust temperatures. Since emission from a
160 K blackbody peaks at 18 $\micron$ and these HMCs have gas temperatures
of 50-200 K, the mid-infrared is a good wavelength regime to try to detect
these types of sources.

I present here sub-arcsecond 18 $\micron$ images of the sites of three hot
molecular cores. All three sites are certified massive star
forming regions due to the presence of UC HII regions. All sites contain
water masers that are isolated from centimeter radio emission sources.
Furthermore, all three sites have been found to contain ammonia emission
coincident with the water maser locations, as observed through
high-resolution molecular line imaging. 

The 18 $\micron$ observations of these HMCs were performed at the Gemini
North and W.M. Keck II telescopes using OSCIR. The reasons for using such
large telescopes are two-fold: 1) to separate the mid-infrared emission of
the HMC from that of any other nearby sources, like extended UC HII regions;
and 2) to exploit the higher sensitivity of the larger telescopes so that
detections are made more readily, or so that stringent upper limits can be
placed on the mid-infrared emission coming from those sources not detected.

\section{Observations}

The University of Florida mid-infrared imager and spectrometer OSCIR was
used for all observations. OSCIR employs a Rockwell 128$\times $128 pixel
Si:As BIB (blocked impurity band) detector. All observations were taken
through the International Halley Watch 18 $\micron$ ({\it IHW18}) filter,
which has an effective central wavelength at 18.06 $\micron$, and a filter
width of 1.7 $\micron$. Background subtraction was achieved during the
observations via the standard chop-nod technique.

While at the Keck II 10-m telescope, OSCIR\ had a pixel scale of 0\farcs0616
pixel$^{-1}$. During the observations point-spread function stars were
observed, yielding an estimate of the effective resolution of the
observations of 0\farcs41. At the Gemini North 8-m telescope, OSCIR had a
pixel scale of 0\farcs084 pixel$^{-1}$ and a effective resolution of 0\farcs%
63.

There were no detections of mid-infrared sources at the locations of 2 of
the 3 hot molecular cores.

\section{The Detection}

\subsection{G29.96-0.02}

\begin{figure}[t]
\plotone{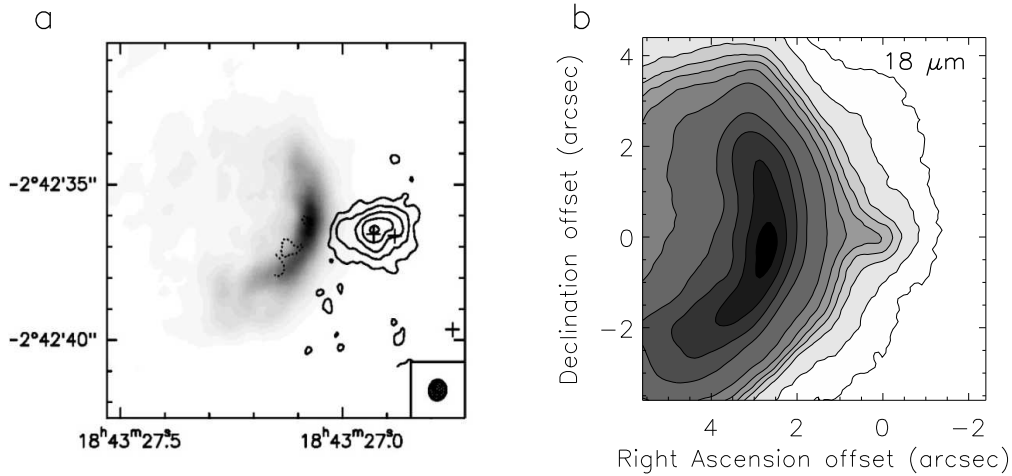}
\caption{Observations of G29.96-0.02. (a) A figure from Cesaroni et al.
(1998). The UC HII region at 1.3 cm (grayscale) and ammonia line emission
(contours). Water maser are plotted as crosses. (b) The 18 $\micron$ image
of the same region shown as filled contours. The HMC can be seen as a
``bump'' on the UC HII region to the west. }
\end{figure}

In the region of G29.96-0.02 lies a cometary shaped ultracompact HII region
around a spectral type O7 star (Watson \& Hanson 1997). In Figure 1a (from
Cesaroni et al. 1998) the grayscale shows the UC HII region at 1.3 cm and
the crosses mark the positions of water masers. These water masers are
offset and do not seem to be associated with the UC HII region. Instead they
appear to be associated with a source seen only in ammonia emission
(contours). It was claimed by Cesaroni et al. (1994) that this source seen
in ammonia emission may be a hot molecular core. Figure 1b shows the 18 $%
\micron$ image from Gemini North of the same region (grayscale with contours
for emphasis). The UC HII region appears more extended in the mid-infrared, but the
HMC can be clearly seen as a ``bump'' on the western side of the UC HII
region.

The HMC in G29.96-0.02 is difficult to see in the mid-infrared because it is
still partially embedded in the extended emission from the nearby UC HII
region. However, a two-dimensional polynomial surface was fit to the this
background emission (excluding a 2$\arcsec$ by 2$\arcsec$ region around the
HMC), isolating the emission from the HMC alone. The three panels in Figure
2 show, from left to right, a portion of the original 18 $\micron$ image
from Gemini containing the HMC, the background polynomial fit of 5th order
in x and y, and the HMC in the residual frame.

\begin{figure}[t]
\plotone{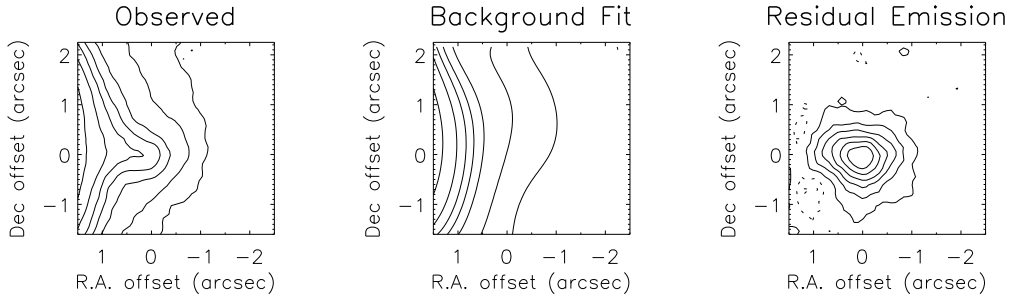}
\caption{Isolating the HMC emission in G29.96-0.02. The three panels show,
from left to right, a portion of the original 18 $\micron$ image from Gemini
containing the HMC, the background polynomial fit of 5th order in x and y,
and the HMC in the residual frame. The contours shown are -16, -8, 8, 25,
42, 59, 76, and 93\% of the peak flux density of 1.9 Jy arsec$^{-2}$.}
\end{figure}

Figure 3a shows a grayscale image of the 1.3 cm continuum observations of
Cesaroni et al. (1998). The overlaying contours are from the 18 $\micron$
mid-infrared data presented in Figure 1. Good morphological agreement for the UC HII region is
found between the two wavelengths, adding credibility to the astrometry. The
large, thin cross marks the position of the ammonia peak of the hot core
from Cesaroni et al. (1998). The small, thick cross marks the peak in the
mid-infrared dust emission from the hot core. Figure 3b shows a close up
view centered on the mid-infrared peak of the hot core. Apparent in this
figure are the offset peaks of the mid-infrared (thick contours) and ammonia
emission (thin contours), possibly due to optical depth effects, or more
likely showing there are two different embedded objects here (De Buizer et
al. 2002). Filled black circles mark the locations of the water masers from
Hofner and Churchwell (1996), and the filled triangle marks the location of
the methanol maser group from Minier, Conway, \& Booth (2001).

A full discussion of the detection of the HMC in G29.96-0.02 can be found 
in De Buizer et al. (2002).

\begin{figure}[t]
\plotone{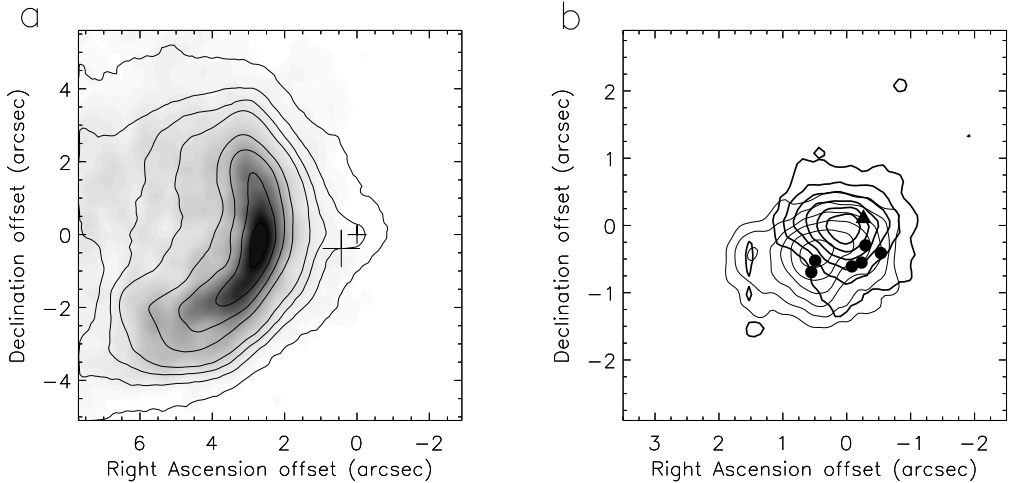}
\caption{Comparing the HMC seen in ammonia and at 18 $\micron$. (a) An
overlay of 18 $\micron$ emission (contours) on the 1.3 cm emission of
Cesaroni et al. (1998) showing the similarity in UC HII region morphology.
The large, thin cross marks the position of the ammonia peak of the hot core
and the small, thick cross marks the peak in the mid-infrared dust emission.
(b) A close up view centered on the mid-infrared peak of the hot core. There
is an apparent offset between the peaks of the mid-infrared (thick contours) and ammonia
emission (thin contours). Filled black circles mark the locations of the
water masers from Hofner and Churchwell (1996), and the filled triangle
marks the location of the methanol maser group from Minier, Conway, \& Booth
(2001).}
\end{figure}

\section{The Non-Detections}

\subsection{G19.61-0.23}

G19.61-0.23 is an extremely complex region of massive star formation. There
are several (at least six) UC HII regions that can be seen in both radio
continuum emission (Garay et al. 1998) and in the mid-infrared. Figure 4a
shows contours of the whole region at 18 $\micron$ from NASA's InfraRed
Telescope Facility 3-m telescope (De Buizer 2000). The stars denote the
locations of three clusters of water masers. The middle of the three water
maser clusters was found to be coincident with a mid-infrared source seen
only at 18 $\micron$ at a detection level of 2.5-$\sigma $ (230 mJy) by De
Buizer (2000). Since such a low detection level is by no means conclusive,
this site required further observation. 

The box in Figure 4a delineates the field of view in Figure 4b taken from
Gemini North. In this figure, contours of ammonia emission (dark gray lines) from
Garay et al. (1998) are overlaid, along with the water masers (crosses) from
Forster and Caswell (1989).  The very bright UC HII region seen in the
mid-infrared is not coincident with the masers, but the masers {\it are}
associated with the ammonia emission. We do not detect a source at any of the
three maser cluster locations at a 3-$\sigma $ upper limit of 12 mJy at 18 
$\micron$, including the middle maser cluster. The weak mid-infrared
emission seen by De Buizer (2000) was apparently noise. 

\begin{figure}[t]
\plotfiddle{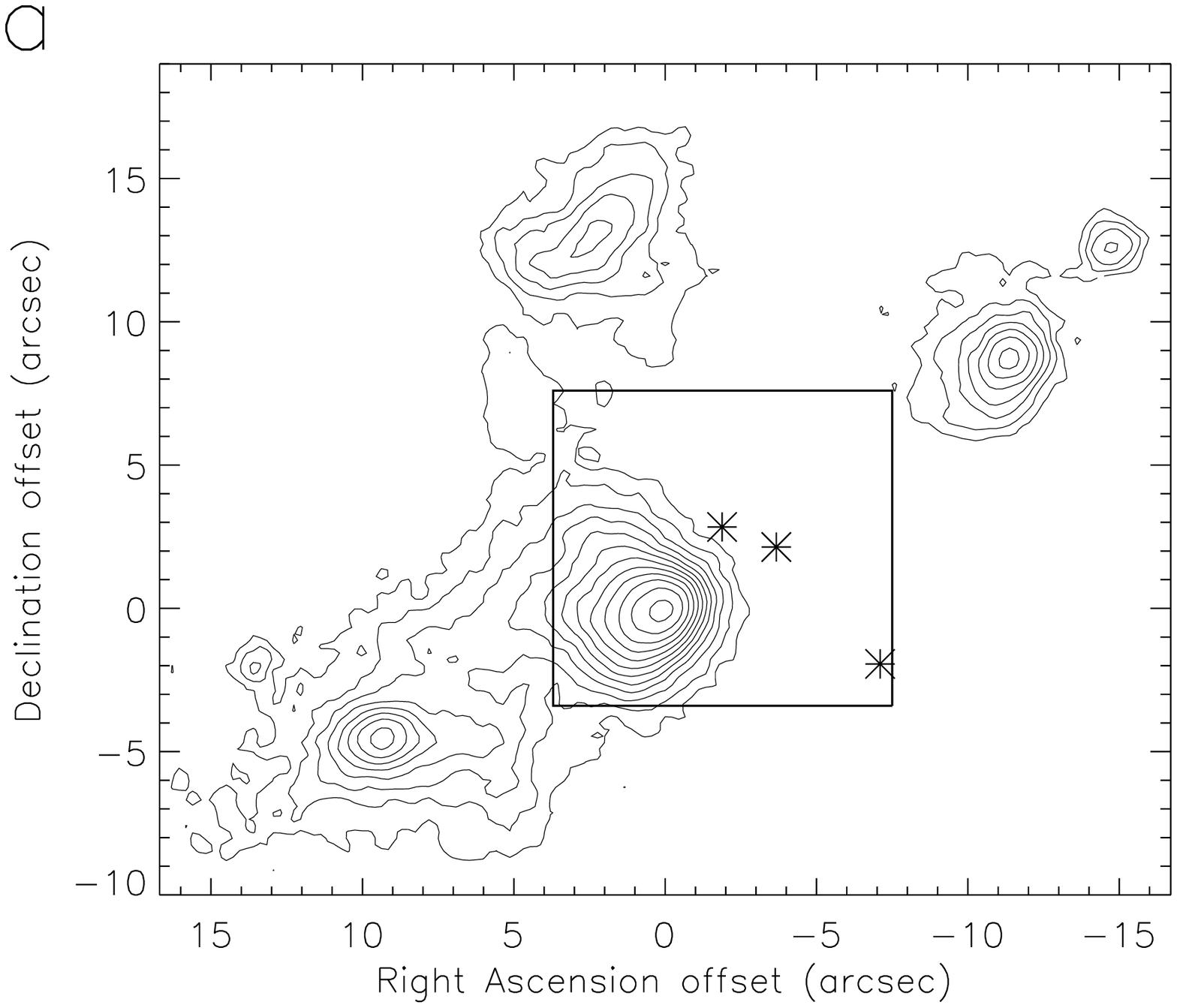}{1.1in}{0}{40}{40}{-200}{-160} %
\plotfiddle{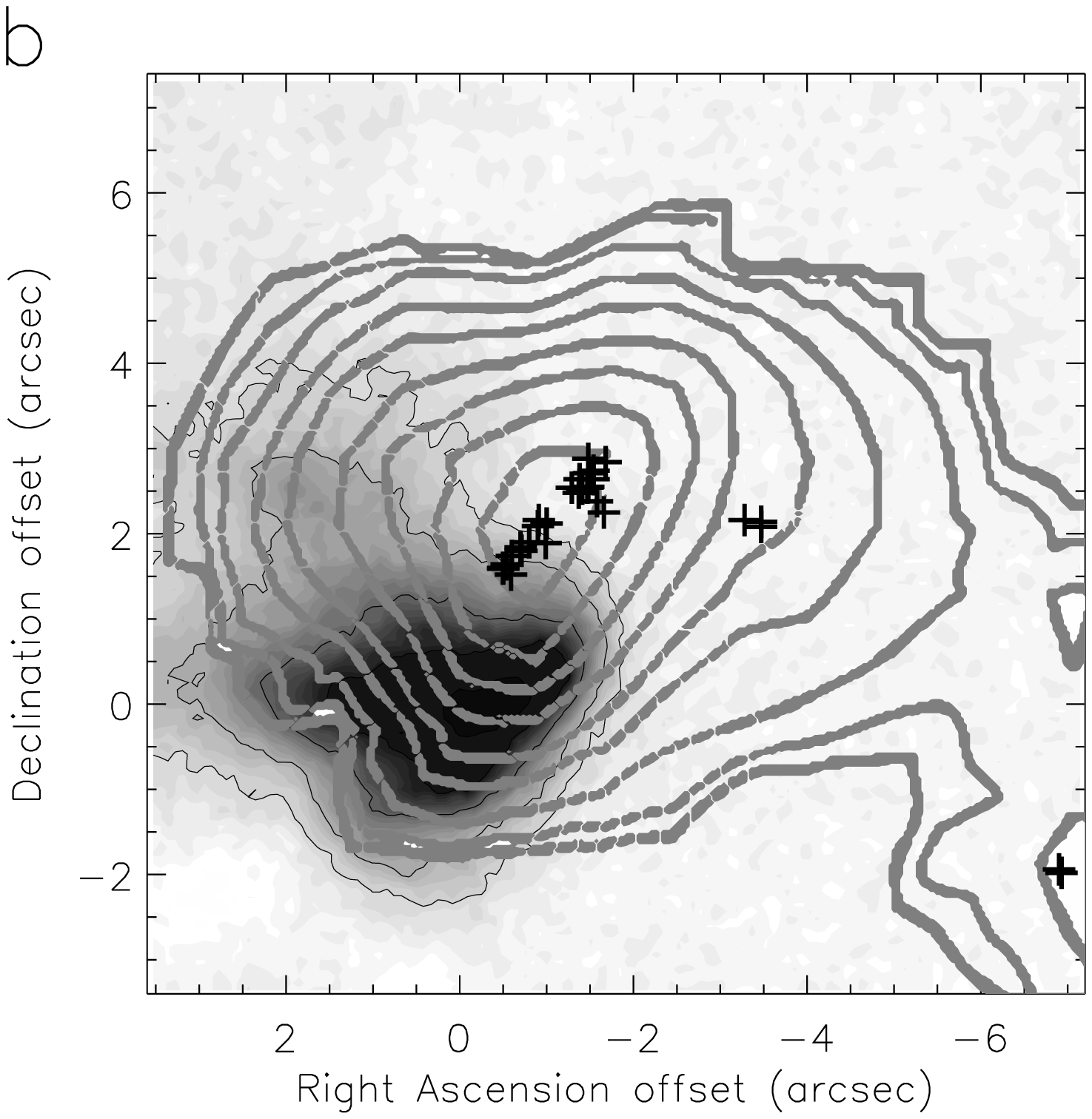}{1.1in}{0}{40}{40}{4}{-67}
\caption{G19.61-0.23. (a) Large-scale 18 $\micron$ emission from De Buizer
(2000). The stars denote the locations of three clusters of water masers, and
the box delineates the field of view of (b). (b) The 18 $\micron$ emission as
seen from Gemini North (grayscale and thin black contours). Overlaid are
contours of ammonia emission (dark gray lines) from Garay et al. (1998) and water
masers (crosses) from Forster and Caswell (1989). We do not detect a source
at any of the three maser cluster locations.}
\end{figure}

\subsection{G34.26+0.15}

Like G29.96-0.02, this site has many masers out in front of the apex of a
cometary shaped UC HII region. However, the masers are more spread out, and
some do appear to be associated with the sharp UC HII boundary. Figure 5a is
from Gaume, Fey, \& Claussen (1994) and shows 2 cm contours of the UC HII
region and two nearby compact sources, as well as the water (crosses) and OH
(circles) masers. In Figure 5b, the light gray contour lines show the location of
ammonia emission from Keto et al. (1992) which is to the east of this UC HII
region and coincident with many masers in the region. Interestingly, Keto et
al. (1992) present a 12.5 $\micron$\ image of this site which shows a faint
(70 mJy) mid-infrared source in the area of the water masers and ammonia
emission. The grayscale image underneath these ammonia contours in Figure 5b
is the 18 $\micron$ image from Gemini North (with thin black
contours for emphasis). No source was detectable at the location of this
ammonia clump at a 3-$\sigma $ upper limit of 20 mJy at 18 $\micron$.
This site was also imaged at 10 $\micron$ by De Buizer (2000) with no
detection at this maser/ammonia location. Therefore, the validity of the source seen
by Keto et al. (1992) is questionable. 

\begin{figure}[t]
\plotone{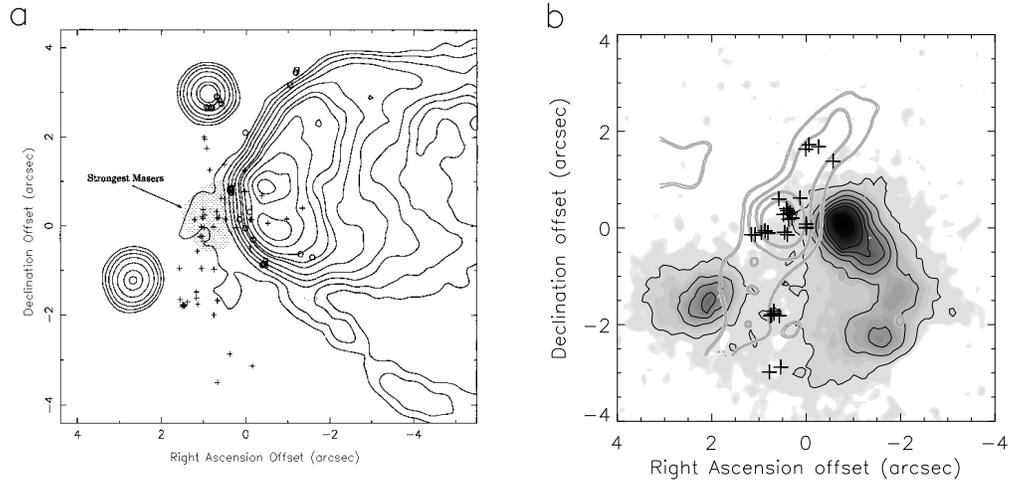}
\caption{G34.26+0.15. (a) A figure from Gaume, Fey, \& Claussen (1994)
showing the 2 cm contours of the UC HII region and two nearby compact
sources, as well as the water masers (crosses). (b) The 18 $\micron$ image from
Gemini North shown as grayscale with thin black contours. The light gray
contours are the ammonia emission from Keto et al. (1992) with the water
masers in the region plotted as crosses. No HMC\ was detected.}
\end{figure}

\section{Conclusions}

Three well-observed hot molecular cores were imaged at 18 $\micron$,
G29.96-0.02, G19.61-0.23, and G34.26+0.15. Only the hot molecular core in
G29.96-0.02 resulted in a detection. This was a surprise considering the
previous ``detections'' of HMCs in the sites of both G19.61-0.23 and
G34.26+0.15. This small research program has shown that the mid-infrared can
be useful in observing and studying some HMCs. It has also shown that the
resolution and sensitivity afforded by the 8 to 10-m class telescopes are
important to perform this research effectively.


\begin{references}

\reference{}Cesaroni, R., Churchwell, E., Hofner, P., Walmsley, C.M., \& Kurtz, S. 1994, \aap, 288, 903
\reference{}Cesaroni, R., Hofner, P., Walmsley, C. M., \& Churchwell, E. 1998, \aap, 331, 709 
\reference{}De Buizer, J. M 2000, Thesis, University of Florida (Available electronically at \\
           http://www.ctio.noao.edu/$\sim$debuizer)
\reference{}De Buizer, J. M., Watson, A. M., Radomski, J. T., Pi\'{n}a, R. K., \& Telesco, C. M. 2002 \apjl, 564, L101
\reference{}Forster, J. R., \& Caswell, J. L. 1989, \aap, 213, 339
\reference{}Garay, G., Moran, J. M., Rodr\'{i}guez, L. F., \& Reid, M. J. 1998, \apj, 492, 635 
\reference{}Gaume, R. A., Fey, A. L., \& Claussen, M. J. 1994, \apj, 432, 648
\reference{}Hofner, P., \& Churchwell, E. 1996, \aap Suppl., 120, 283
\reference{}Keto, E., Proctor, D., Ball, R., Arens, J., \& Jernigan, G. 1992, \apjl, 401, L113   
\reference{}Minier, V., Conway, J. E., \& Booth, R. S. 2001, \aap, 369, 278
\reference{}Watson, A. M., \& Hanson, M. M. 1997, \apjl, 490, L165

\end{references}
\end{document}